# Valley optoelectronics based on meta-waveguide photodetectors

Chi Li[1]†, Kaijian Xing[2]†, Wenhao Zhai[3], Luca Sortino[4], Andreas Tittl[4], Igor Aharonovich[5,6], Michael S. Fuhrer[1], Kenji Watanabe[7], Takashi Taniguchi[8], Qingdong Ou[2]*, Zhaogang Dong[3,9,10]*, Stefan A. Maier[1,11]*, Haoran Ren[1]*

[1]School of Physics and Astronomy, Monash University, Melbourne, VIC, 3800, Australia.

[2]Macao Institute of Materials Science and Engineering (MIMSE), Faculty of Innovation Engineering, Zhuhai MUST Science and Technology Research Institute, Macau University of Science and Technology, Taipa, Macao 999078, China.

[3]Institute of Materials Research and Engineering (IMRE), Agency for Science, Technology and Research (A*STAR), Singapore 138634, Republic of Singapore.

[4]Chair in Hybrid Nanosystems, Faculty of Physics, Ludwig-Maximilians-Universität München, 80539, Munich, Germany.

[5]School of Mathematical and Physical Sciences, University of Technology Sydney, Ultimo, New South Wales 2007, Australia.

[6]ARC Centre of Excellence for Transformative Meta-Optical Systems, University of Technology Sydney, Ultimo, New South Wales 2007, Australia

[7]Research Center for Electronic and Optical Materials, National Institute for Materials Science, 1-1 Namiki, Tsukuba 305-0044, Japan

[8]Research Center for Materials Nanoarchitectonics, National Institute for Materials Science, 1-1 Namiki, Tsukuba 305-0044, Japan

[9]Quantum Innovation Centre (Q.InC), Agency for Science Technology and Research (A*STAR), Singapore 138634, Republic of Singapore

[10]Singapore University of Technology and Design, Singapore 487372, Republic of Singapore

[11]Department of Physics, Imperial College London, London SW7 2AZ, U.K.

Correspondence emails: qdou@must.edu.mo, zhaogang_dong@sutd.edu.sg, stefan.maier@monash.edu, haoran.ren@monash.edu

† These authors contributed equally to this work.

**Abstract:**

In transition metal dichalcogenides, the valley degree of freedom directly couples valley-polarised excitons—excited by circularly polarised light—to valley-dependent chiral photons, enabling ultrafast light-driven valleytronics. However, achieving fully integrated valley optoelectronics—incorporating on-chip generation, selective routing, and electrical readout of valley-dependent chiral photons—remains an unresolved challenge. We present a valley-driven hybrid nanophotonic-optoelectronic circuit that integrates chirality-selective meta-waveguide photodetectors with transition metal dichalcogenides. At room temperature, our purposely designed meta-waveguide device generates near-unity valley-dependent chiral photons in the second harmonic generation from an encapsulated tungsten disulfide monolayer and selectively couples them to unidirectional waveguide modes, achieving an exceptional polarisation selectivity of 0.97. These valley-dependent waveguide modes were subsequently detected by atomically thin few-layer tungsten diselenide photodetectors, exclusively responsive to the above-bandgap upconverted photons, thereby enabling all-on-chip processing of valley-multiplexed images. Our demonstration bridges a critical gap in lightwave valleytronics, paving the way for compact, scalable valley information processing and fostering the development of light-based valleytronic quantum technologies.

**Main:**

The valley degree of freedom has emerged as a transformative platform for ultralow-power information processing (*1-3*) and quantum computing (*4-6*). Valleytronics leverages asymmetric electron populations in momentum-distinguished valleys—degenerate conduction band minima—to encode information. By integrating on-chip transport and electrical readout of valley-polarised currents, this approach enables robust, scalable systems for valley information processing. At its core, valley polarisation arises from broken spatial inversion symmetry or time-reversal symmetry in a material, lifting the energetic degeneracy between the K and K' valleys in the first Brillouin zone (*7-9*). This valley polarisation can be dynamically controlled via external stimuli such as magnetic fields (*10*), electric fields (*8*), and strain engineering (*11*), enabling electronic valley devices and Hall transistors in diverse material platforms, including graphene (*12*), graphene/hexagonal boron nitride (hBN) superlattices (*13*), bilayer graphene (*3, 8*), and diamond (*14*).

Transition metal dichalcogenides (TMDCs) are prime candidates for lightwave valleytronics, offering optical control over valley polarisation at an ultrafast speed. Specifically, K and K′ valleys can be selectively addressed via circularly polarised light in monolayers (*15-18*), heterostructures (*19-23*) or topological optical fields in bulk systems (*9*). Monolayer TMDCs, in particular, have significant advantages as direct bandgap semiconductors, enabling efficient coupling of valley-polarised excitons to valley-dependent chiral photons. At room temperature, this addresses key challenges of short exciton lifetimes and low charge mobility in the transport of valley-polarised currents (*2, 24*), thereby opening the possibility of efficient, high-speed, and long-range transport of valley-dependent photons.

Photoluminescence from monolayer TMDCs, governed by the optical Stark effect, exhibits limited valley polarisation at room temperature (*17, 25-27*). Alternatively, exciton-enhanced second-harmonic generation (SHG), driven by the Bloch-Siegert effect (*28*), achieves near-unity valley polarisation under ambient conditions (*29-32*). Despite progress in the generation (*9, 33, 34*), free-space steering (*35, 36*), and on-chip routing (*37-42*) of valley-dependent chiral photons, previous nanophotonic platforms are not fully integrated, relying on bulky optical detection of valley information in the far field. On-chip optoelectronics, which integrates optical and electronic components on a single chip and features both ultrafast optical processing and efficient electrical readout, holds great promise (*43*). However, the realisation of fully integrated light-driven valley optoelectronic circuits—combining on-chip generation, selective transport, and electrical readout of valley-dependent chiral photons—remains an unresolved challenge.

Here we demonstrate a valley-driven hybrid nanophotonic-optoelectronic circuit integrated with TMDCs for on-chip generation, selective routing, and electrical readout of valley-dependent chiral photons at room temperature (Fig. 1a). Our valley meta-waveguide photodetector (VMP) combines a hBN-encapsulated tungsten disulphide ($WS_2$) monolayer (Fig. 1b)—achieving room-temperature near-unity valley-dependent chiral SHG photons (Fig. 1c)—with a chirality-selective metasurface that routes SHG photons to left- or right-propagating dielectric waveguide modes (polarisation selectivity: 0.97). These valley-dependent waveguide modes were subsequently detected by few-layer tungsten diselenide ($WSe_2$) photodetectors, responsive only to above-bandgap SHG photons (Fig. 1d), enabling all-on-chip optical processing and digital electrical readout of valley-multiplexed images (Fig. 1e). Our demonstrated valley-driven hybrid nanophotonic-optoelectronic circuit has opened the possibility of all-on-chip processing of the valley degree of freedom in TMDCs at room temperature. This advancement will unlock the full potential of light-based valleytronics for both classical and quantum photonic applications.

**Results:**

A VMP device integrates a meta-waveguide with photodetectors. Figure 2a illustrates the meta-waveguide model for circular polarisation-selective light coupling and directional routing. The model comprises two key components: a chirality-selective metasurface made of amorphous silicon (a-Si) and a silicon nitride ($Si_3N_4$) waveguide. The metasurface separates photons with opposite circular polarisation by imparting opposite linear momentum, enabling unidirectional coupling to the waveguide mode. To achieve this, the unit cell and meta-atom dimensions—pitch ($p$), length ($l$), width ($w$) and height ($h$)—were meticulously optimised through parametric sweeps. Leveraging a geometric phase metasurface, precise wavefront phase control was achieved by rotating the individual meta-atoms by an angle ($\varphi$). Details of the methodology and sweep results are provided in the Supporting Information (Fig. S1). In 3D electromagnetic wave modelling, a set of coplanar electric dipoles served as the input source, with the dipole to metasurface distance denoted as $d$. Waveguide-coupled light fields were measured at a distance greater than 30$\lambda$ from the metasurface centre, which are labelled as "L" and "R" for left- and right-end of the waveguide, respectively. Using our optimised meta-atom parameters, unidirectional waveguide coupling was achieved at a wavelength of 620 nm (corresponding to the $WS_2$ bandgap) under left ($\sigma^-$) and right ($\sigma^+$) circularly polarised light (Fig. 2b).

A histogram of waveguide-coupled field intensity versus incident wavelength under left ($\sigma^-$) circularly polarised illumination is presented in Fig. 2c. At the desired wavelength, the intensity difference between left ($I_L$, red) and right ($I_R$, blue) sides under $\sigma^-$ illumination reaches approximately two orders of magnitude. The $\sigma^+$ illumination case leads to a symmetric optical response similar to the $\sigma^-$ case, and thus we omitted it for brevity, as will be the case below unless specifically noted. In practical scenarios, the electrical dipoles from a transferred $WS_2$ monolayer are likely positioned at a non-zero distance ($d$) from the metasurface. To explore this, we investigated how $d$ affects coupling intensity and absolute polarisation selectivity ($|(I_L-I_R)/(I_L+I_R)|$).

It shows that the absolute polarisation selectivity (marked by green circles) remains above 0.92 with high intensity imbalance ($|I_L-I_R|$) (marked by black squares) when $d < 80$ nm (Fig. 2d). Beyond this threshold, the coupling selectivity decreases sharply due to the reduced light-metasurface interaction cross-section with increasing distance.

The meta-waveguide was fabricated using standard electron-beam lithography followed by dry plasma etching (see Methods and Fig. S2 in the Supporting Information). A scanning electron microscopy (SEM) image of the fabricated sample is presented in Fig. 2e. Two grating couplers, designed for a wavelength of 620 nm, were fabricated at both ends of the waveguide to enable far-field pre-characterisation by outcoupling the waveguide mode to free space. The fabricated structures exhibit excellent agreement with the designed geometric parameters, as highlighted by the zoomed-in SEM images in Fig. 2e. The fabricated sample was experimentally characterised by illuminating the metasurface with a focused laser beam and detecting scattered light from the outcoupling gratings across various wavelengths and polarisations (Fig. 2f and 2g). Our fabricated meta-waveguide showed strong circular polarisation selectivity at the target wavelength of 620 nm. This selectivity decreases as the wavelength was reduced from 620 nm to 550 nm but increases again at wavelengths below 550 nm, which may be attributed to the wavelength sensitivity of the used grating couplers. The measured intensity difference between the left and right outcoupling gratings was nearly one order of magnitude. Our measured polarisation selectivity, assessed by rotating a quarter-wave plate, reached a maximum value of ~ 0.97 when the polarisation aligned with the metasurface, closely matching the simulation value of ~ 0.997. Related images are available in Supporting Information (Fig. S3). This high polarisation selectivity of our meta-waveguide device could facilitate subsequent on-chip detection of valley-dependent chiral photons.

The generation of valley-dependent chiral photons utilises excitonic processes with an optical pump of circularly polarised light at room temperature. We mechanically exfoliated a monolayer

WS$_2$ flake and protected it with a thin hBN flake. The flake stacking configuration and excitation scheme are illustrated in Fig. 3a, and the configuration of optical setup is described in Fig. S4. Initially, the sample was excited at 594 nm (near the excitonic transition at ~620 nm) and polarisation resolved PL spectra were recorded for σ⁻ (top) and σ⁺ (bottom) excitations (Fig. S5). After polarisation filtering, PL emission with the same polarisation as the excitation exhibited stronger intensity, with a modest valley polarisation of approximately 10%, consistent with other reports at room temperature (*25, 44*). In contrast, valley-dependent SHG exhibited near-unity valley polarisation in a nonlinear process. We show that the SHG intensity increases as the excitation wavelength decreases from 1400 nm, reaching the maximum when it approaches twice the WS$_2$ excitonic transition (Fig. 3b). A further reduction in the excitation wavelength results in a sharp decline in the SHG signal, accompanied by a secondary PL peak (Fig. S6), which is likely attributed from two-photon PL. Opposite chirality was observed between the pump and SHG polarisation states, highlighting the distinct valley selection rules governing linear and nonlinear processes. Furthermore, the intensity of valley-dependent chiral SHG photons exhibits a quadratic dependence on the excitation power at a pump wavelength of 1240 nm, with the power function fitting yielding a factor of 2.05 (Fig. 3b inset).

Subsequently, we examined unidirectional waveguide routing of valley-dependent chiral photons by transferring an exfoliated WS$_2$ monolayer (encapsulated within hBN flakes) onto the metasurface region of the meta-waveguide (Fig. 3c). Gold pads were deposited around the waveguide to facilitate photodetector fabrication and streamline the transfer process. The WS$_2$ was pumped with a focused femtosecond laser at 1240 nm under σ⁻ and σ⁺ polarisation states, respectively. The resulting valley-dependent chiral SHG photons were selectively routed in opposite directions, as evidenced by prominent scattering from one end of the waveguide, captured in the far-field optical images (Fig. 3d). Consistent with laser demonstration (Fig. S3), σ⁻ (σ⁺) excitation generates $\sigma^+$ ($\sigma^-$) SHG photons, which were routed to the right (left) ends of the waveguide, respectively.

We completed the VMP device by fabricating two few-layer $WSe_2$-based photodetectors positioned at the ends of the meta-waveguide. The TMDC flakes used in the two photodetectors (PD-1 and PD-2) and the metasurface region are outlined with different colours and styles for clarity (Fig. 4a), with their stacking orders schematically illustrated in Fig 4b. The fabricated photodetectors demonstrated a robust photoresponse at the target wavelength of 620 nm, as shown by the I-V plot (Fig. S7) and their ON/OFF response under varying incident powers (Fig. S8). At an excitation power of 10 nW, measured at the back focal plane of the illumination objective lens, our $WSe_2$ photodetector successfully detected photocurrent without the need for signal amplification.

We then directly measured the valley-dependent chiral photons using the integrated photodetectors by illuminating the $WS_2$ monolayer with $\sigma^+$ (Fig. 4c) and $\sigma^-$ (Fig. 4d) polarised laser, respectively. Notably, our photodetectors respond exclusively to above-bandgap SHG photons, while remaining unresponsive to the excitation lasers. Both photodetectors displayed distinct source-drain currents ($I_{sd}$) with clear ON/OFF behaviour, exhibiting high currents for their respective target polarisation. Specifically, $\sigma^+$ ($\sigma^-$) incidence generates $\sigma^-$ ($\sigma^+$) valley-dependent chiral photons, routing to the left- (right-)side photodetectors and producing dominant currents in PD-1 (-2), shown as red (blue) lines, respectively. The green square signals represent laser's "ON" and "OFF" states, controlled by an optical chopper operating at 20 Hz. The weak cross-polarisation $I_{sd}$ signals are comparable, plotted as blue line in PD-1 and red line in PD-2, reflecting a small routing crosstalk in these two channels. The photocurrent in PD-1 does not fully decay to the dark current level within one ON-OFF cycle, likely due to electron trapping caused by imperfect contact and flake transfer (*45, 46*). This issue could potentially be mitigated through improved device fabrication techniques (*47, 48*).

The polarisation sensitivity of the photodetectors was further verified by measuring net valley photocurrents while rotating a quarter-wave plate. The recorded photocurrents exhibit a distinct sinusoidal trend, with variations of approximately 340 pA for both PD-1 and -2 (Fig. 4e). This leads to photocurrent polarisation selectivity of 0.63 (0.56) for PD-1 (-2).

To showcase our device's potential in valley optoelectronics and on-chip information processing, we encoded two images of "Kangaroo" and "Koala" (256 by 256 pixels) onto σ− (σ+) valley-dependent chiral photons, corresponding to σ+ (σ−) light incidence, respectively (Fig. 4f). This valley-multiplexed image data (Fig. 4g) was simultaneously sent through the VMP for all-on-chip processing of valley information. PD-1 and PD-2 successfully decoded the σ− (σ+) valley-dependent chiral photons and their associated "Kangaroo" (Fig. 4h) and "Koala" (Fig. 4i) images. Strong current signals (bright image pixels) were recorded from the photodetectors for the correct valley states, whereas weak current signals (dark image pixels) corresponded to the incorrect valley states. Specifically, PD-1 displays a distinct "kangaroo" image, despite with slightly lower image contrast resulting from moderate photocurrent variations (Fig. 4e). We note that these dark pixels can be entirely eliminated by applying a pre-characterised photocurrent threshold during digital data processing. As a result, our demonstration validates that the developed VMP optoelectronic circuit successfully facilitates all-on-chip processing of valley-dependent chiral photons at room temperature.

**Conclusions:**

We have demonstrated a valley-driven hybrid nanophotonic-optoelectronic circuit enabled by the integration of ultracompact meta-waveguide photodetectors with TMDCs. This circuit achieves on-chip generation, selective routing, and electrical readout of valley-dependent chiral photons. At room temperature, near-unity valley-dependent chiral SHG was generated from a hBN-encapsulated WS$_2$ monolayer. Our chirality-selective meta-waveguide, exhibiting a high polarisation selectivity of 0.97 in far-field characterisation, incorporates few-layer WSe$_2$ photodetectors for the selective electrical readout of valley-dependent upconverted chiral photons, achieving an on-chip photocurrent polarisation selectivity of 0.63. Our developed ultracompact valley-driven hybrid nanophotonic-optoelectronic circuit has enabled the first demonstration of all-on-chip optical processing and electrical readout of valley-multiplexed image information. We believe our demonstration bridges a critical gap in valley optoelectronics, paving the way for compact, scalable processing of valley-dependent photon emission and advancing the development of light-based valleytronic architectures.

**Acknowledgments:** This work was performed in part at the Melbourne Centre for Nanofabrication (MCN) in the Victorian Node of the Australian National Fabrication Facility (ANFF). The authors thank H. D. for the assistance on LabVIEW program. We are grateful for TMDC crystals supported by J. H. The authors thank S. W. for the discussion of the photocurrent measurements.

**Funding**: The authors acknowledge below funding supports.

H. R.: Australian Research Council grant (DE220101085, DP220102152);

S. A. M.: Australian Research Council grant DP220102152, Lee Lucas Chair in Physics;

I. A.: Australian Research Council (CE200100010, FT220100053);

Q. O.: National Natural Science Foundation of China (52402166), the Science and Technology Development Fund, Macau SAR (0065/2023/AFJ, 0116/2022/A3), and the Australian Research Council (DE220100154);

M. S. F.: Australian Research Council grant DP200101345;

Z. D.: Agency for Science, Technology and Research (A*STAR) under its MTC IRG (Project No. M22K2c0088), National Research Foundation via Grant No. NRF-CRP30-2023-0003 and SUTD Kickstarter Initiative (SKI) grant with the award No. SKI 2021_06_05;

A. T.: Funded by the European Union (ERC, METANEXT, 101078018). Views and opinions expressed are however those of the author(s) only and do not necessarily reflect those of the European Union or the European Research Council Executive Agency. Neither the European Union nor the granting authority can be held responsible for them. Also funded by the Deutsche Forschungsgemeinschaft (DFG, German Research Foundation) under grant numbers EXC 2089/1–390776260 (Germany's Excellence Strategy) and TI 1063/1 (Emmy Noether Program);

L.S.: Humboldt Research Fellowship from the Alexander von Humboldt Foundation.

K. W. and T. T. acknowledge support from the JSPS KAKENHI (Grant Numbers 21H05233 and 23H02052) , the CREST (JPMJCR24A5), JST and World Premier International Research Center Initiative (WPI), MEXT, Japan.

**Author contributions:** C. L., Q. O., Z. D., and H. R. conceived the project. C. L. and H. R. modelled the meta-waveguides. C. L., W. Z. and Z. D. fabricated the meta-waveguide samples. Z. D. contributed to the scanning electron microscope. K. W. and T. T. provided hBN flakes. K. X. designed and performed the experiments regarding 2D material and photodetector integration. K. X, C. L. and H. R. characterised the meta-waveguide photodetectors. C. L. processed data with advice from K. X., H. R., L. S. and Q. O.. C. L., K. X., and H. R. wrote the first draft of the manuscript. A. T., I. A., S. A. M., M. S. F. contributed to discussion of the results. H. R. supervised the project.

**Competing interests:** The authors declare that they have no competing interests.

**Data and materials availability:** All data are available in the main text or the supplementary materials.

**Supplementary Materials**

Methods

Supplementary Text

Figs. S1 to S8

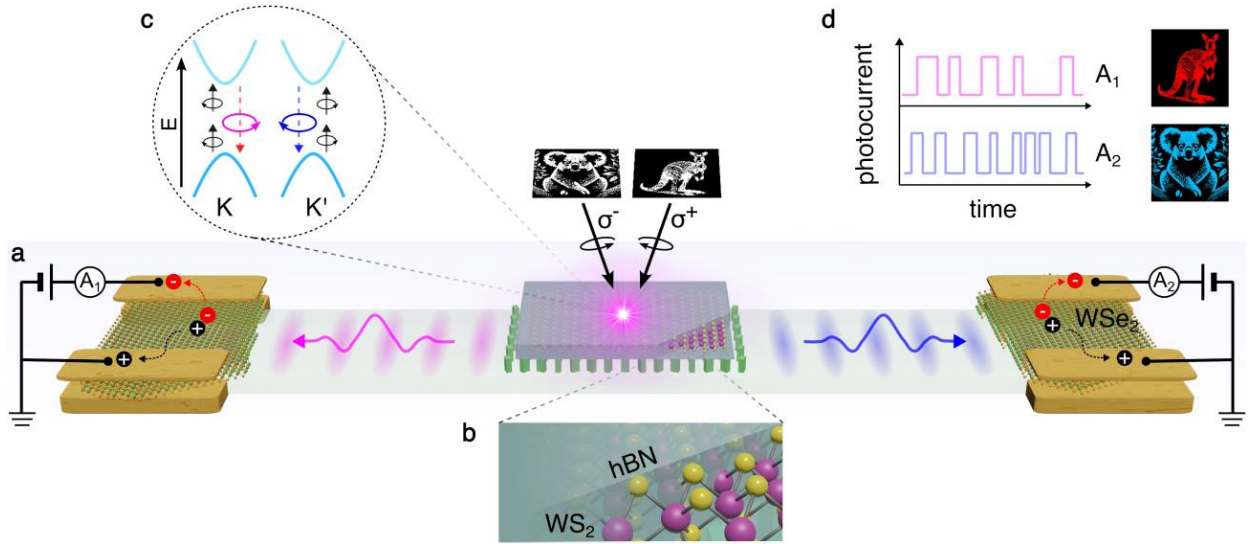

**Fig. 1: Schematic of a valley-driven hybrid nanophotonic-optoelectronic circuit based on TMDCs.** (**a**) Principle of all-on-chip generation, selective routing, and electrical detection of valley-dependent chiral photons in a meta-waveguide photodetector device. Valley-dependent chiral SHG photons are generated from monolayer $WS_2$ atop meta-waveguide, which are coupled to unidirectional waveguide modes and further excite electron-hole pairs separation in few-layer $WSe_2$-based photodetectors under electrical bias. (**b**) Zoomed view of a hBN-encapsulated $WS_2$ monolayer. (**c**) Band diagram illustration of nonlinear valley selection rule governing the generation of valley-dependent chiral SHG photons. (**d**) On-chip processing of valley polarisation-multiplexed image information. Schematic of photodetector readouts of time-sequential photocurrents and their reconstructed valley polarisation-encoded Koala (σ-) and Kangaroo (σ+) images.

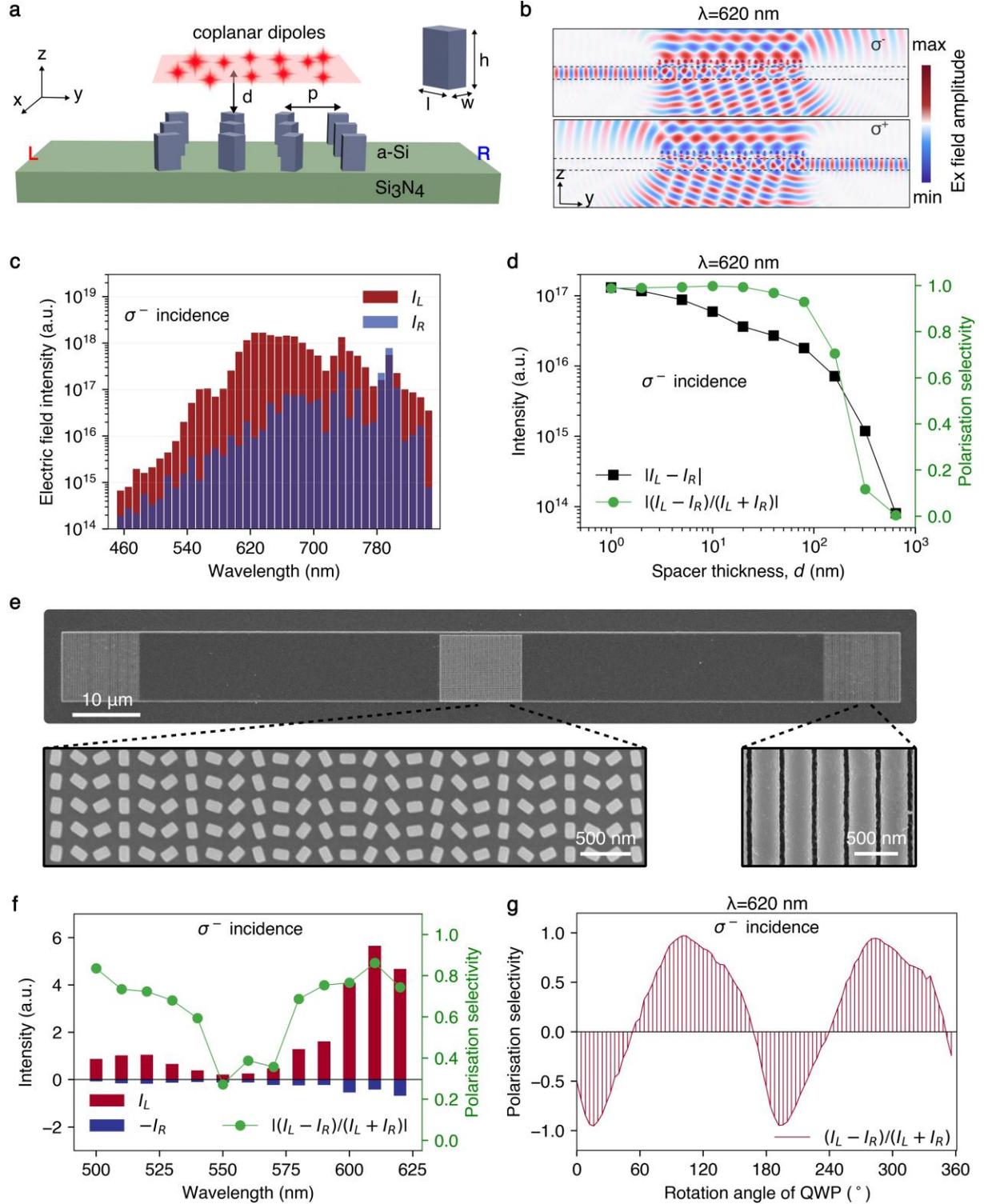

**Fig. 2: Design, fabrication, and far-field characterisation of a chirality-selective meta-waveguide.** (**a**) 3D configuration of a proposed meta-waveguide model, with a-Si metasurface addressed on a $Si_3N_4$ waveguide sitting on a quartz substrate (not shown in the schematic). A set of coplanar electric dipoles are placed above the meta-atoms for illumination. The meta-atoms are defined by length $l$, width $w$, height $h$, pitch $p$, and rotation angle $\varphi$. (**b**) Electric field distributions

for the σ- (top) and σ+ (bottom) polarisation at a wavelength of 620 nm, respectively. (**c**) Numerically calculated electric field intensity collected from the waveguide ends "L" and "R". (**d**) Dipole to meta-atom distance "*d*" dependent light coupling and selective routing. (**e**) SEM image of a fabricated meta-waveguide sample. Metasurface and grating coupler regions are zoomed in for clarity, respectively. (**f**) Experimentally measured selective routing of circularly polarised light in the far-field for different wavelengths. (**g**) Measured polarisation selectivity at the designed wavelength of 620 nm.

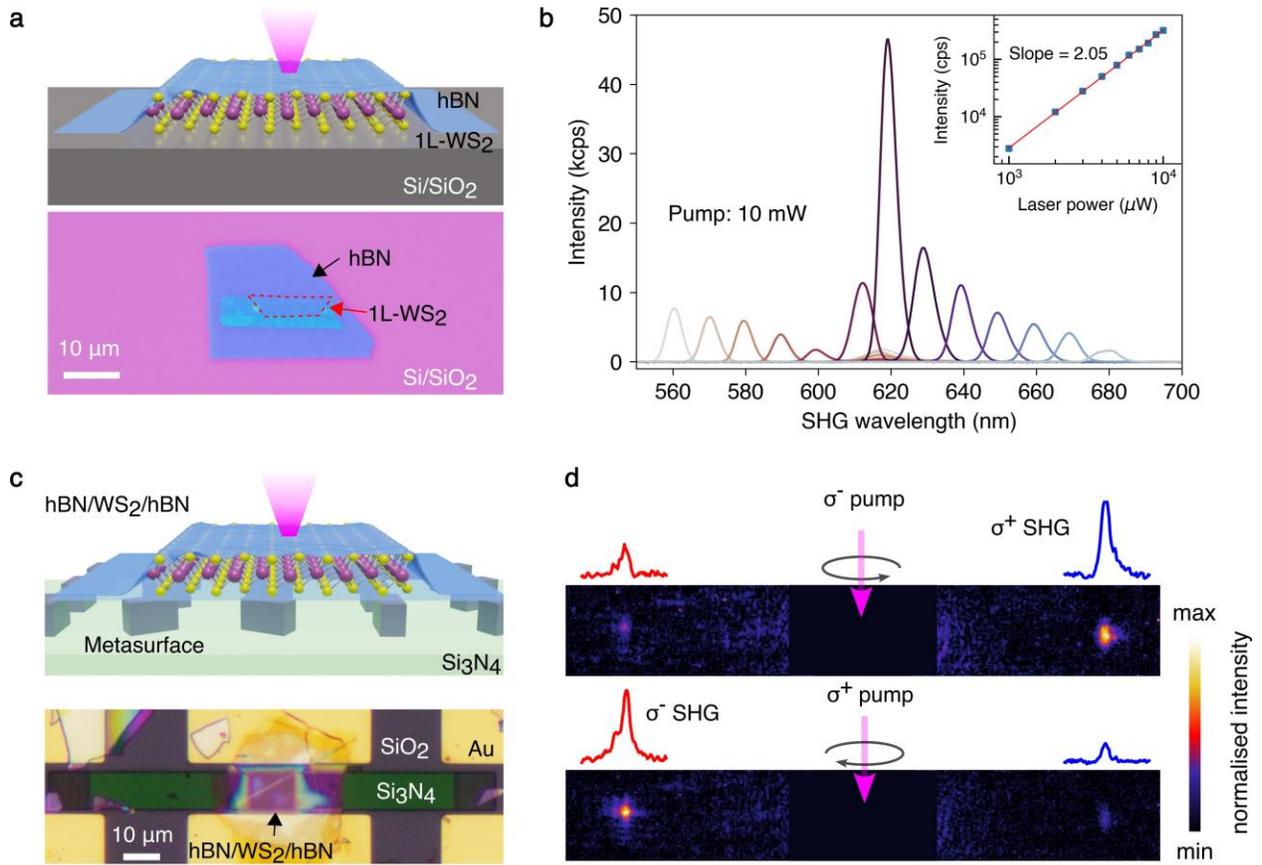

**Fig. 3: Generation of valley-dependent chiral SHG photons and their selective routing by a meta-waveguide device.** (**a**) Schematic and optical image of a thin hBN-encapsulated WS$_2$ monolayer on a flat oxidised silicon substrate. (**b**) Valley-dependent SHG spectra of monolayer WS$_2$ under cross-polarisation analysis ($\sigma^-$ pump and $\sigma^+$ detection) at different wavelengths. Inset, SHG intensity versus pump powers at the wavelength of 1240 nm, showing a clear quadratic dependence. (**c**) Schematic and optical image of a valley-driven meta-waveguide device featuring a thin hBN-encapsulated WS$_2$ monolayer, transferred onto the metasurface region. (**d**) Far-field characterisation of the selective routing of valley-dependent chiral SHG photons by the meta-waveguide device in (c), performed at a wavelength of 1240 nm. The metasurface region at the laser beam spot was removed from the images to highlight the evident SHG intensities.

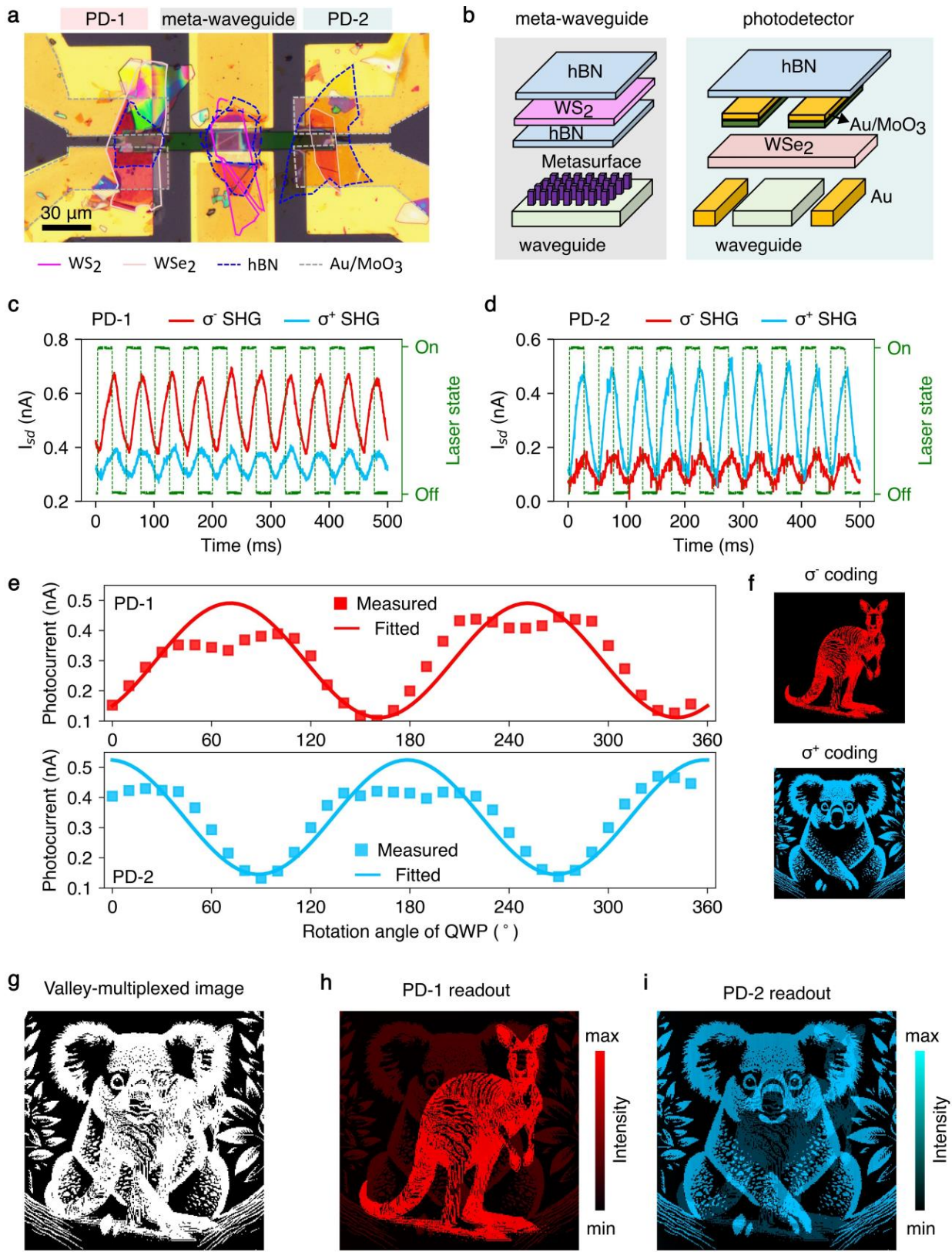

**Fig. 4: Photocurrent readout of valley-dependent chiral SHG photons and on-chip processing of valley information.** (**a**) An optical image of the fabricated VMP device. The metasurface and photodetector regions are labelled in different colours, each representing different materials, to enhance clarity. (**b**) Schematic illustration of material stacking approach. (**c** and **d**) The ON/OFF photocurrent responses of photodetectors PD-1 (c) and PD-2 (d) under different valley polarisation states. (**e**) Polarisation selectivity of the VMP device in response to different incident polarisation states controlled via a quarter wave plate. (**f**) Target images of "Kangaroo" and "Koala" encoded with $\sigma^-$ and $\sigma^+$ valley polarisations, respectively. (**g-i**) On-chip processing of the valley-multiplexed image (g) yields valley-dependent readouts, with predominant image responses of "Kangaroo" (h) and "Koala" (i) reconstructed from the left and right photodetectors, respectively.